\documentclass{article}
\usepackage{spconf,amsmath,graphicx}
\usepackage{caption}

\usepackage{setspace}

\title{A Foundation Model for Music Informatics}
\name{Minz Won, Yun-Ning Hung, and Duc Le}
\address{SAMI, ByteDance, San Jose, CA, USA}
\begin{document}
%
\maketitle
\begin{abstract}
This paper investigates foundation models tailored for music informatics, a domain currently challenged by the scarcity of labeled data and generalization issues. To this end, we conduct an in-depth comparative study among various foundation model variants, examining key determinants such as
model architectures, tokenization methods, temporal resolution, data, and model scalability. This research aims to bridge the existing knowledge gap by elucidating how these individual factors contribute to the success of foundation models in music informatics. Employing a careful evaluation framework, we assess the performance of these models across diverse downstream tasks in music information retrieval, with a particular focus on token-level and sequence-level classification. Our results reveal that our model demonstrates robust performance, surpassing existing models in specific key metrics.
These findings contribute to the understanding of self-supervised learning in music informatics and pave the way for developing more effective and versatile foundation models in the field. A pretrained version of our model is publicly available to foster reproducibility and future research.

\end{abstract}

\section{Introduction}
\label{sec:introduction}
A foundation model refers to any pretrained machine learning model capable of being adapted to a broad spectrum of downstream tasks~\cite{Bommasani2021FoundationModels}. By taking advantage of its self-supervised nature, AI researchers have been able to scale up foundation models with enormous data, resulting in versatile representation that can generalize to diverse downstream tasks in multiple domains, such as natural language processing~\cite{devlin2018bert,brown2020language}, computer vision~\cite{dosovitskiy2020image,arnab2021vivit}, speech recognition~\cite{hsu2021hubert,chiu2022self}, and multimodal research~\cite{radford2021learning}.

The potential of foundation models is not only limited to the aforementioned domains. In the field of music information retrieval (MIR), researchers have consistently faced challenges in scaling up their models, mainly due to the lack of labeled data. The task of annotating music data is labor-intensive and often demands specialized domain knowledge, which hinders large-scale data collection efforts. To circumvent these limitations, researchers have employed semi-supervised learning~\cite{won2021semi,kum2020semi} or transfer learning~\cite{van2014transfer,choi2017transfer} schemes, which leverage knowledge from larger labeled datasets. Nevertheless, these approaches often encounter generalization issues, particularly when the downstream tasks involve information not represented in the supervised data~\cite{mccallum2022supervised}. Also, scalability remains constrained by the availability of labeled data. This has led to a growing interest in developing self-supervised foundation models specifically tailored for music informatics research.

Although research on foundation models in music informatics is still in its nascent stages, several pioneering works have made notable contributions.  CLMR~\cite{spijkervet2021contrastive} and MULE~\cite{mccallum2022supervised} have employed a simple contrastive learning framework~\cite{chen2020simple} to capture sequence-level music representation that summarizes the entire sequence, rather than preserving information at every time step. While their performance lags behind that of supervised methods, they have nonetheless demonstrated the potential for self-supervised models to generalize across multiple music tagging tasks. Another innovative study~\cite{castellon2021codified} has revealed that language models pretrained on tokenized representations, such as Jukebox~\cite{dhariwal2020jukebox}, can serve as robust foundation models for various downstream MIR tasks. The authors discussed that the generative model could learn richer representations than conventional tagging models. More recently, MERT~\cite{li2023mert} has adapted HuBERT~\cite{hsu2021hubert}, a successful speech recognition foundation model, to formulate a framework explicitly tailored for music representation, proving its generalizability across an array of sequence-level classification tasks. 

However, despite these advances, it remains unclear how individual factors, such as model architectures, tokenization methods, temporal resolution, data, and model scalability, contribute to the success of foundation models in music informatics. This knowledge gap underscores the need for a comprehensive investigation, which is the focus of our research. In this work, we compare a new self-supervised learning approach from speech recognition (i.e., BEST-RQ~\cite{chiu2022self}) with MERT~\cite{li2023mert} through a meticulous evaluation of both token-level and sequence-level downstream classification tasks. We find that our model exhibits robust performance across a range of MIR tasks and outperforms existing models in specific contexts. A pretrained version of our model is available online~\footnote{https://github.com/minzwon/musicfm} to facilitate reproducible research.





\begin{figure}[htb]
\centering
\centerline{\includegraphics[width=6.9cm]{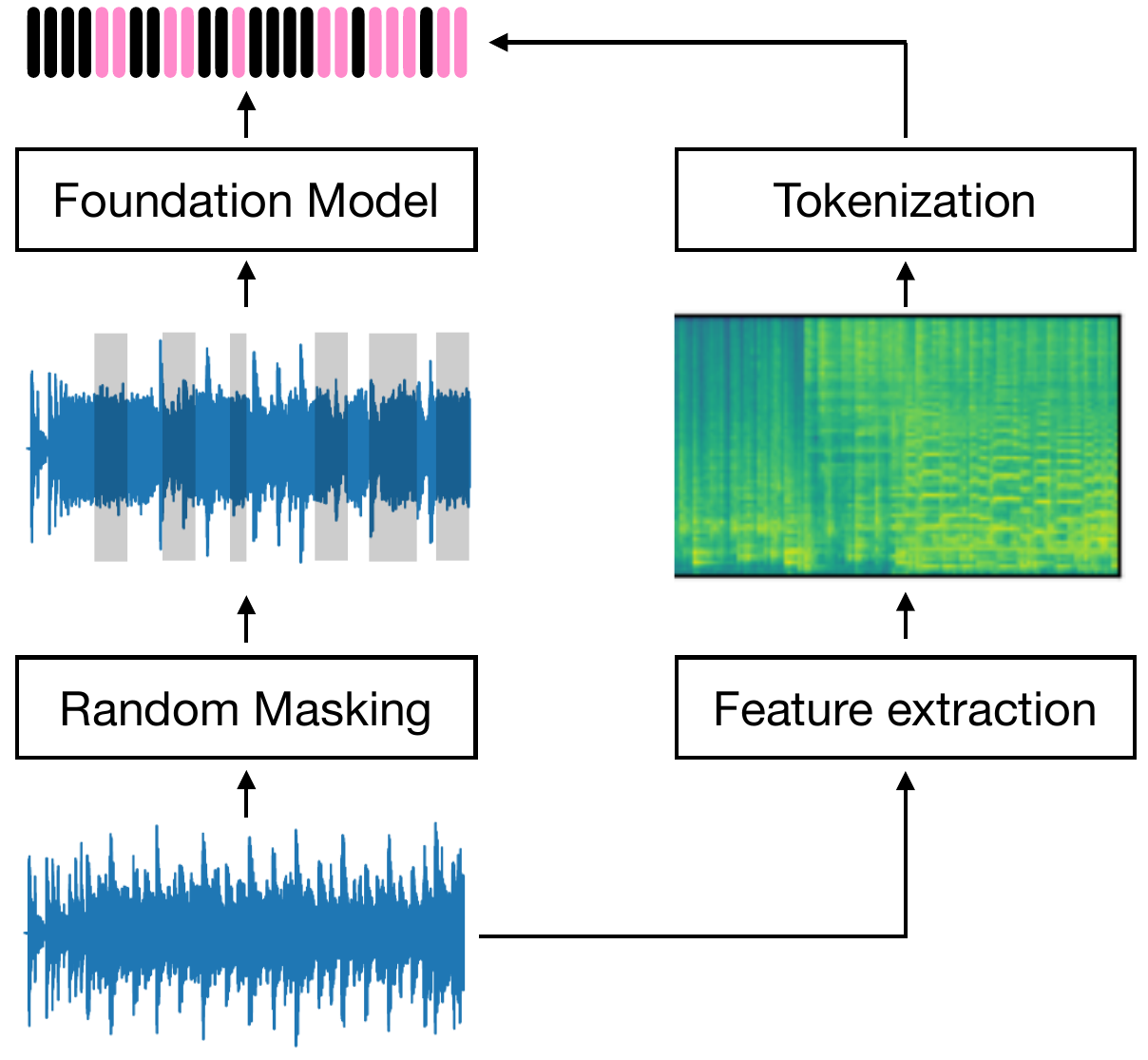}}
%
\caption{Masked token modeling of audio representation.}
\label{fig:model}
\end{figure}

\section{Models}
\label{sec:models}

This section outlines the key concepts of foundation models that are employed in this study. Given that the objective of our research is to examine the various factors that contribute to the successful development of foundation models, we leverage existing models and training methodologies from previous works~\cite{chiu2022self, li2023mert}. It is important to note that none of the methods described in this section represent our original contributions; they serve as the basis for our investigations into optimizing foundation models for music informatics.

\subsection{Masked token modeling}
Alongside generative models such as GPT-3~\cite{brown2020language}, masked token modeling techniques, such as BERT~\cite{devlin2018bert}, have shown robust performance as foundation models across a variety of sequential data types, particularly in speech audio~\cite{baevski2020wav2vec,zhang2020pushing,hsu2021hubert,chiu2022self}. In the masked token modeling paradigm, the objective is to predict tokens that have been deliberately masked within a sequence. As illustrated in Figure~\ref{fig:model}, portions of the input audio are randomly masked with noise. The foundation model is then tasked with predicting these intentionally masked or omitted tokens (highlighted in pink). This approach enables the foundation model to learn contextualized semantics, which is useful for various downstream tasks. While various model architectures can serve as the backbone for a foundation model, transformer variants~\cite{baevski2020wav2vec,zhang2020pushing,hsu2021hubert,chiu2022self} are most commonly used due to their exceptional sequence modeling capabilities. Specifically, the BERT-style encoder~\cite{hsu2021hubert} has been incorporated into the MERT~\cite{li2023mert}, and the Conformer~\cite{zhang2020pushing} has been utilized in BEST-RQ~\cite{chiu2022self}, which are core backbones of our experiments.
Masked token modeling can be formalized as a self-supervised classification task using the tokenization methods (right side of Figure~\ref{fig:model}) that will be introduced in the following subsection.

\subsection{Tokenization}
To cast music sequence modeling within the masked token modeling framework, it is essential to convert short audio segments into discrete tokens, a process known as tokenization. While various tokenization methods are available, our baseline, MERT~\cite{li2023mert}, utilizes k-means clustering~\cite{hsu2021hubert} and residual vector quantization (RVQ)~\cite{defossez2022high}. The authors employ k-means clustering on log-mel spectra to capture timbral characteristics and chroma features for encoding harmonic attributes. The resulting representations are then tokenized according to their corresponding feature clusters. 

However, most conventional tokenization methods, including k-means clustering and RVQ, necessitate a separate training phase for representation learning. This additional step can introduce complexities and create dependencies that may impact the foundation model's overall performance. To address these challenges, recent work (BEST-RQ~\cite{chiu2022self}) proposed a tokenization method employing a random projection quantizer to bypass the need for a trainable representation learning phase. This approach has two random components: random projection and random codebook lookup, both of which obviate the need for training. Within this scheme, a \( d \)-dimensional input vector \( x \) is mapped to an \( h \)-dimensional latent space via random projection \( R \). The closest index from a randomly initialized \( n \times h \) codebook \( C \) is then selected as the feature-representing token. The tokenized representation \( \tau \) can be formalized as:

\begin{equation}
\tau = \arg\min_i \left| \lVert c_i \rVert_2 - \lVert Rx \rVert_2 \right|,
\end{equation}
where \( R \) is an \( h \times d \) matrix for random projection, \( c_i \) represents the \( i \)-th vector in random codebook \( C \), and \( \lVert \cdot \rVert_2\) denotes the \( l_2 \)-norm. The projection matrix \( R \) is initialized with Xavier initialization, while the codebook \( C \) uses standard normal distribution. Log-mel spectra serve as the only feature vectors and are normalized to have zero mean and unit variance. This normalization step is particularly crucial when employing a non-trainable random projection, as codebook utilization can be very low without it.


\section{Experiments}
\label{sec:experiments}

\subsection{Datasets}
We utilize two distinct datasets to train our foundation models. The first dataset consists of 160k hours of in-house music data, designed to align with the size of the data used to train MERT. The second dataset is the Free Music Archive (FMA) dataset~\cite{fma_dataset}, which comprises 8k hours of Creative Commons-licensed music audio. All audio files from both datasets have been preprocessed to a standard format of 24kHz mono audio.


\subsection{Evaluation}
Previous studies~\cite{spijkervet2021contrastive,castellon2021codified,mccallum2022supervised,li2023mert} have largely focused on evaluating music foundation models through sequence-level classification tasks, such as genre classification, emotion recognition, key detection, and music tagging. However, many applications in music information retrieval necessitate predictions at individual time steps. Given this context, we propose that robust foundation models should demonstrate strong capabilities in token-level classification tasks, such as beat tracking and chord recognition. To test this hypothesis, our research evaluates foundation models across five distinct tasks: beat tracking, chord recognition, structure analysis (all token-level classification tasks), as well as key detection and music tagging (both sequence-level classification tasks).

Consistent with prior work~\cite{castellon2021codified}, we employ a probing model on top of the foundation model for our evaluation. This probing model is structured as a shallow neural network with a single 512-dimensional hidden layer and an output layer. Since our goal is to scrutinize the pretrained representations, the foundation model remains frozen during this probing stage. In the case of sequence-level classification, we use an average pooling layer to aggregate the representations across time before initiating linear probing.

\vspace{0.2cm}

\noindent\textbf{Beat/downbeat Tracking} 
aims to predict the timestamps of each beat and the position of the first beat in each bar. In accordance with previous studies~\cite{Hung2022ModelingBA}, our model generates frame-level probabilities of beats, downbeats, and non-beat events every 50 ms. To decode downbeat timestamps, we employ a dynamic Bayesian network (DBN) implemented in \texttt{madmom}~\cite{Bck2016JointBA} for post-processing.
Harmonix Set~\cite{nieto2019harmonix} is used for training, while GTZAN~\cite{Marchand2015GTZANRhythmET} is used as a test set. We use F-measure implemented in \texttt{mir\_eval}~\cite{Raffel2014MIR} for evaluation.

\vspace{0.2cm}

\noindent\textbf{Chord Recognition}
is a challenging MIR task due to the intricate harmonic relationships within music compositions. In this work, we focus solely on major and minor chords. The model outputs frame-level probabilities of 25 classes, which include the 12 pitches of major and minor chords, along with one category denoted as ``none,'' and does so at every 125 ms.
We use HookTheory~\cite{Kum2020SemisupervisedLU} for training, with 2000 songs selected for testing. The evaluation metric is major/minor weighted accuracy in \texttt{mir\_eval}~\cite{Raffel2014MIR}.

\vspace{0.2cm}

\noindent\textbf{Structure Analysis}
aims to segment a music recording into distinct, non-overlapping sections and predict the functional label for each segment, such as `verse' and `chorus.' Following previous settings~\cite{Wang2022ToCA}, the probing model has two classifiers designed to predict frame-level probabilities of seven functional classes and boundaries. Each frame has a resolution of 200 ms. We select 150 pieces from Harmonix Set~\cite{nieto2019harmonix} for testing, and the rest of Harmonix Set is used for training. We use the F-measure of hit rate at 0.5 seconds (\emph{HR.5F}) to evaluate boundary and frame-wise accuracy to evaluate functional labels. Both metrics are computed using \texttt{mir\_eval}~\cite{Raffel2014MIR} package.

\vspace{0.2cm}

\noindent\textbf{Key Detection}
aims to predict the tonal scale and pitch relation across the entire songs. The model has to output frame-level probabilities of 25 classes (12 major and 12 minor keys plus one ``none'') per 2 seconds. We use HookTheory~\cite{Kum2020SemisupervisedLU} for training. Giantsteps~\cite{Knees2015TwoDS} is used as test set. The evaluation metric is refined accuracy (weighted accuracy) implemented by \texttt{mir\_eval}~\cite{Raffel2014MIR}, with error tolerance that gives partial credits to reasonable errors.

\vspace{0.2cm}

\noindent\textbf{Music tagging}
subsumes various music classification tasks~\cite{won2021music}, such as genre, mood, instrument, and language classification. Since any musical attribute can be music tags, music tagging serves as a comprehensive downstream task to gauge a model's versatility. We employed the widely-used MagnaTagATune dataset~\cite{law2009evaluation}, consisting of popular 50 tags. We maintained the same data splits as those employed in a previous study~\cite{won2020evaluation}. Evaluation metrics are the mean average precision (mAP) and the area under the receiver operating characteristic curve (ROC-AUC).



\subsection{Foundation models}
We implemented a self-supervised learning approach, as detailed in Section~\ref{sec:models}, which closely follows the methodology outlined in related work~\cite{chiu2022self}. This approach takes advantage of random quantization, eliminating the requirement for separate representation learning. In line with the referenced paper, our codebook consists of 8192 16-dimensional vectors. We applied a 400ms window with a 60\% probability mask, utilizing only the masked segment for cross-entropy loss optimization. FM8 in Table~\ref{table:results} follows the exact same setup of the previous work in the speech domain, BEST-RQ~\cite{chiu2022self}, except for the data.

In this study, we aim to address the fundamental question of how to build an advanced foundation model. To achieve this objective, we carefully scrutinize each component to assess their individual impacts. Our investigation encompasses an exploration of random quantization, an in-depth examination of various architectures, including a BERT-style encoder from HuBERT~\cite{hsu2021hubert} and Conformer\cite{zhang2020pushing}, while considering different model size configurations, as well as an investigation into varying temporal resolutions (i.e., the number of tokens per second), diverse input lengths for pretraining, and two distinct datasets. The summary of various model configurations can be found in Table~\ref{table:results}.

We utilized the Adam optimizer~\cite{kingma2014adam} with a learning rate of 0.0001. Learning rate warm-up over 30,000 steps played a critical role in our training process, especially when using \texttt{float16}. To enhance training efficiency, we employed deepspeed~\cite{rasley2020deepspeed}, flash attention~\cite{dao2022flashattention}, and mixed precision techniques. All models were trained using eight A100-80GB GPUs for two weeks.

\begin{table*}[t]
    \centering
    \captionsetup[table*]{skip=5pt}
    \setlength{\abovecaptionskip}{8pt}
    \caption{Foundation model variants and their respective downstream metrics. FM1* corresponds to MERT~\cite{li2023mert}, while FM8** mirrors the BEST-RQ~\cite{chiu2022self} but with the distinction that it was trained using music data.}
    \label{table:results}
    \renewcommand{\arraystretch}{0.90}
    \resizebox{1.0\textwidth}{!}{%
    \begin{tabular}{|l|c|c|c|c|c|c|c|c|c|c|c|c|c|c|}
        \hline
        \multicolumn{7}{|c|}{Foundation model}  & \multicolumn{2}{c|}{Beat} & Chord & \multicolumn{2}{c|}{Structure} & Key & \multicolumn{2}{c|}{Tagging}\\ \hline
        Index & Encoder & Size & Hz & Input & Token & Data & Beat F1 & Downbeat F1 & Acc & Acc & HR.5 & Acc & mAP & ROC\\ \hline
        FM1\textasteriskcentered & BERT & 330M & 75Hz & 5s & K-means & In-house & 0.858 & 0.722 & 0.574 & 0.578 & 0.626 & 0.645 & 0.4499 & 0.9167 \\ \hline
        FM2 & BERT & 330M & 75Hz & 5s & Random & In-house & 0.856 & 0.669 & 0.636 & 0.588 & 0.490 & 0.636 & 0.4185 & 0.9013 \\ \hline
        FM3 & BERT & 330M & 75Hz & 30s & Random & In-house & 0.855 & 0.703 & 0.651 & 0.640 & 0.641 & 0.662 & 0.4226 & 0.9000 \\ \hline
        FM4 & Conformer & 330M & 75Hz & 5s & Random & In-house & 0.863 & 0.771 & 0.643 & 0.619 & 0.534 & 0.660 & 0.4589 & 0.9161 \\ \hline
        FM5 & Conformer & 330M & 75Hz & 30s & Random & In-house & 0.865 & 0.780 & 0.702 & 0.715 & 0.710 & 0.670 & 0.4821 & 0.9208 \\ \hline
        FM6 & Conformer & 330M & 50Hz & 30s & Random & In-house & 0.864 & 0.802 & 0.690 & 0.710 & 0.698 & 0.671 & 0.4816 & 0.9204 \\ \hline
        FM7 & Conformer & 330M & 25Hz & 30s & Random & In-house & 0.868 & \textbf{0.804} & \textbf{0.714} & \textbf{0.726} & 0.699 & 0.674 & \textbf{0.4883} & \textbf{0.9235} \\ \hline
        FM8\textasteriskcentered\textasteriskcentered & Conformer & 660M & 25Hz & 30s & Random & In-house & 0.866 & 0.800 & 0.689 & 0.722 & \textbf{0.716} & 0.649 & 0.4790 & 0.9216 \\ \hline
        FM9 & Conformer & 330M & 25Hz & 30s & Random & FMA & 0.868 & 0.767 & 0.675 & 0.664 & 0.631 & 0.674 & 0.4726 & 0.9167 \\ \hline \hline
        FM7-finetune & Conformer & 330M & 25Hz & 30s & Random & In-house & 0.865 & 0.803 & \textbf{0.807} & \textbf{0.742} & 0.740 & 0.715 & 0.4809 & 0.9198 \\ \hline
        FM8-finetune & Conformer & 660M & 25Hz & 30s & Random & In-house & 0.874 & \textbf{0.810} & 0.800 & 0.739 & \textbf{0.744} & 0.700 & 0.4735 & 0.9192 \\ \hline
        FM9-finetune & Conformer & 330M & 25Hz & 30s & Random & FMA & 0.861 & 0.785 & 0.784 & 0.718 & 0.737 & 0.690 & 0.4695 & 0.9168 \\ \hline \hline
        \multicolumn{7}{|l|}{State-of-the-art \cite{Hung2022ModelingBA, jonggwon_park_2019_3527886, taejun2023allinone, Wang2022ToCA, filip_korzeniowski_2018_1492399, alonso2023pre}} & \textbf{0.887} & 0.756 & 0.762 & 0.723 & 0.660 & \textbf{0.746} & 0.470 & 0.913 \\ \hline
    \end{tabular}
    }
\end{table*}

\section{Results}
\label{sec:resuts}
\vspace{-2pt}
This section presents our experimental results and findings. Our carefully designed ablation study elucidates the critical factors of developing successful foundation models for music.

\vspace{0.2cm}
\noindent\textbf{Random tokenization}~\cite{chiu2022self} generalizes well to music data. Despite the absence of an additional representation learning stage, it effectively acquires useful representations for various downstream tasks. Notably, even when random tokenization is exclusively applied to mel spectra, it surpasses FM1 (MERT) in the chord recognition task. This achievement is particularly noteworthy given that FM1 leverages an additional auxiliary task involving the constant-Q transform (CQT) reconstruction for capturing harmonic information. 
In the structure analysis task, an initial performance gap exists between FM1 and FM2, attributed to FM2's omission of auxiliary tasks, this gap is effectively bridged by employing longer input sequences (FM3).



\vspace{0.2cm}
\noindent\textbf{Token-level classification} offers a more comprehensive understanding of foundation models. When we solely compare various models using sequence-level classification, such as music tagging, it can be challenging to discern significant differences among them. However, token-level classification tasks, particularly those that demand a longer-term context, such as downbeat tracking and structure analysis, distinctly expose the limitations of models trained on shorter input sequences (FM2 and FM4).

\vspace{0.2cm}
\noindent\textbf{Input length} used during training is critical for capturing long-term contexts. Foundation models pretrained with 5s inputs (FM1, FM2, and FM4) excel in tasks related to timbre, such as music tagging, or tasks that only rely on local contexts, such as beat tracking. However, they exhibit lower performance than models trained with 30s inputs in tasks like downbeat tracking and structure analysis. We believe that advanced foundation models should be capable of modeling both long-term and short-term contexts, making longer inputs a recommended choice.

\vspace{0.2cm}
\noindent\textbf{Temporal resolution} has less impact in our experimental setup. A model with 25Hz temporal resolution (FM7) demonstrated slightly superior performance compared to its 50Hz (FM6) and 75Hz (FM5) counterparts, while demanding less computation due to shorter sequence lengths.

\vspace{0.2cm}
\noindent\textbf{Model architecture} makes a significant difference. Conformer (FM5) consistently outperformed BERT encoder (FM3) for across all downstream tasks. Interestingly, the influence of model size was relatively minimal (FM7 and FM8). However, it's worth noting that larger models might require longer training and more meticulous optimization to fully realize their performance potential.

\vspace{0.2cm}
\noindent\textbf{Data} is undeniably crucial, as in any data-driven approach. A model pretrained with 160k-hour music audio (FM7) showed better performance compared to a model trained on the 8k-hour FMA dataset (FM9). Two factors may contribute to this difference: first, the scalability, and second, the crowd-sourced nature of FMA, which encompasses a considerable amount of noisy data, impacting the model's generalizability.

\vspace{0.2cm}
\noindent\textbf{Fine-tuning} the foundation model further enhances downstream performance. However, we did observe a performance drop in the tagging task, primarily attributed to overfitting.

\section{Conclusion}
\label{sec:conclusion}
\vspace{-8pt}
In this study, we conducted a thorough investigation into foundation model settings, evaluating their effects across five music information retrieval tasks, spanning both token-level and sequence-level classifications. Our experiments revealed six critical factors in the development of foundation models. As a result, our foundation model consistently surpasses its predecessor in all downstream tasks, with a notable performance gap, particularly evident in token-level classification tasks that necessitate long-term context.
We anticipate that this enhanced foundation model will make valuable contributions, not only in classification tasks but also in the realms of multi-modal retrieval and generative models, as it incorporates a richer musical context.

\vfill\pagebreak




\bibliographystyle{IEEEbib}
\small
\setstretch{1.0}

\bibliography{strings,refs}

\end{document}